\documentclass[aps,showpacs]{revtex4}

\usepackage{graphicx}
\usepackage{amsmath}
\usepackage{amssymb}

\newcommand{\be}{\begin{equation}}
\newcommand{\ee}{\end{equation}}
\newcommand{\ba}{\begin{eqnarray}}
\newcommand{\ea}{\end{eqnarray}}

\newcount\bozza \bozza=0
\ifnum\bozza=1
\newdimen\shift \shift=-2truecm
\def\lb#1{%
{\label{#1}\rlap{\kern\shift{$\scriptstyle#1$}}}}
\else\def\lb#1{\label{#1}} \fi

\begin{document}


\title{Magnetic oscillations in planar
systems with the Dirac-like spectrum \\
of quasiparticle excitations II: transport properties}

\author{V.P.~Gusynin$^{1}$}
\email{vgusynin@bitp.kiev.ua}
\author{S.G.~Sharapov$^{2}$}
\email{sharapov@bitp.kiev.ua}
\thanks{Present address: Bogolyubov Institute for Theoretical Physics, Kiev, Ukraine}

\affiliation{$^1$
Bogolyubov Institute for Theoretical Physics,
        Metrologicheskaya Str. 14-b, Kiev, 03143, Ukraine\\
        $^2$Istituto Nazionale per la Fisica della Materia
        (INFM);\\
        Institute for Scientific Interchange, via Settimio Severo 65,
I-10133 Torino, Italy}

\date{\today }

\begin{abstract}
The quantum magnetic oscillations of electrical (Shubnikov de Haas
effect) and thermal conductivities are studied for graphene which
represents a distinctive example of planar systems with a linear,
Dirac-like spectrum of quasiparticle excitations. We show that if
a utmost care was taken to separate electron and phonon
contributions in the thermal conductivity, the oscillations of
electron thermal conductivity, $\kappa(B)$ and Lorenz number,
$L(B)$ would be observable in the low-field (less than a few
Teslas) regime.
\end{abstract}




\maketitle

\section{Introduction}

There is a variety of condensed matters systems that in the first
approximation can be regarded as planar and which in the low
energy sector can be described by the Dirac-like form of the
effective Hamiltonian. The difference in the behavior of the
particles with the usual parabolic spectrum and the linear,
Dirac-like spectrum becomes particularly prominent when an
external quantizing magnetic field $B$ is applied perpendicularly
to the plane.  The energies of Landau levels for free
nonrelativistic electrons are $E_n = e \hbar B/(mc) (n+
\frac{1}{2})$, while for the ``relativistic'' problem 
\be \lb{rel}
E_n = \sqrt{e \hbar v_F^2 B 2 n/c}, 
\ee 
with $n =0,1, \ldots$. Here
$e$ is the electron charge, $m$ is the effective mass of carriers
in the parabolic band, $v_F$ is the Fermi velocity of the system
with linear dispersion, and we wrote $\hbar$ and $c$ explicitly;
in the following sections we set $\hbar = c = k_B =1$, unless
stated explicitly otherwise.

The most important qualitative difference between these two
spectrums is that for the realistic values of the parameters $m$
and $v_F$ rather weak fields $B \sim 10 \,\mbox{Tesla}$  is
sufficient to drive ``relativistic'' systems in the extreme
quantum regime \cite{Abrikosov:1998:PRB} causing such interesting
phenomena as quantum magnetoresistance. Another qualitatively
distinguishable feature of the Dirac-like quasiparticles is an
unusual form of the semiclassical quantization condition for
energy levels in the magnetic field $S(\epsilon) = 2 \pi e
B/(\hbar c) (n+ \gamma) $, where $S$ is the cross-sectional area
of the orbit in $\mathbf{k}$ space, $n$ is a large integer
($n>0$), and $\gamma$ is a constant ($0 \leq \gamma \leq 1$). For the
parabolic dispersion $\gamma =1/2$, which is commonly used in
describing magnetic oscillation (MO) phenomena in metals
\cite{Shoenberg.book}, so that the corresponding Berry's phase,
$\gamma - 1/2$ is zero. However, for the Dirac quasiparticles
Berry's phase is nontrivial \cite{Mikitik:1999:PRL} and $\gamma
=0$, so that the commonly used expressions \cite{Shoenberg.book}
have to be modified accordingly. This was indeed obtained in our
previous paper \cite{Sharapov:2004:PRB}, where we have studied  MO
of the density of states (DOS), thermodynamic potential, and
magnetization in QED$_{2+1}$ with the spectrum (\ref{rel}). We
have also  discussed the underlying condensed matter models and
systems that in the low-energy approximation are reduced to
QED$_{2+1}$ form. Among them is graphite, which is probably the
most promising and convenient for experimental investigation of
the ultraquantum regime, when only a few lowest Landau levels are
occupied. This semimetal was originally studied almost 50
years ago and there was a considerable renewal of the interest in
the electronic properties of this material during the past decade
due to the discovery of novel carbon-based materials such as
carbon nanotubes constructed from wrapped graphene sheets
\cite{Saito:book}. While graphite itself is a three-dimensional
material in which planar sheets of carbon atoms are stacked,
graphene is an individual layer or sheet made from the carbon
atoms.

The simplified QED$_{2+1}$ description is obviously appropriate
for graphene. In Kish graphite the anisotropy of the resistivity
$\rho_c$ along the $c$-axis direction and the basal-plane
resistivity, $\rho_b$ can be $\rho_c/\rho_b \sim 10^2$ and even as
large as $\rho_c/\rho_b \sim 5 \times 10^4$ in  highly oriented
pyrolytic graphite (HOPG) \cite{Kopelevich:2002}, indicating that
the layers are weakly coupled. Therefore in the first
approximation, the QED$_{2+1}$ description with some limitations may
also be used for graphite. Indeed, early theoretical
investigations of graphite show that while low-lying Landau levels
correspond to Eq.~(\ref{rel}), there are other levels related to
the warping of the graphite Fermi surface
\cite{Dresselhaus:1974:PRB}. Very recent measurements of de Haas
van Alphen (dHvA) and Shubnikov de Haas (SdH) oscillations in HOPG
\cite{Luk'yanchuk:2004} (for an earlier literature see
Refs.~\cite{Uji:1998:PB,Wang:2003:PLA}) confirm that among other
carriers in graphite there is a majority holes with 
tw-dimensional (2D) Dirac-like spectrum. The dependences of 
the thermal and Hall conductivities on the applied magnetic 
field in HOPG were studied in Ref.~\cite{Ocana:2003:PRB} and 
more comprehensive data on thermal
conductivity and the deviations from the Wiedemann-Franz (WF) law were
reported in \cite{Ulrich:2004:JLTP}. We mention aslo recent STS
observations \cite{Matsui:2004} of Landau levels at graphite
surfaces. Finally we refer to monocrystalline graphitic films made
by repeated peeling of small mesas of HOPG
\cite{Novoselov:2004:Science}. These films contain only a few
layers of graphene. The SdH oscillations are clearly observed in
this material \cite{Novoselov:2004:Science} and they depend only
on the perpendicular component of the applied magnetic field. This
proves the 2D nature of the material. Moreover, the carrier density
(and even the character of carriers, either electrons or holes) in
this system is controllable by electric-field doping, so that
there are SdH oscillations with varying applied voltage
\cite{Novoselov:2004:c-m}.

The purpose of the present paper is to extend the analysis of our
previous paper \cite{Sharapov:2004:PRB} and study the MO of the
electrical, $\sigma(B)$ (SdH effect) and thermal $\kappa(B)$
conductivities. We demonstrate that since the field dependences of
$\sigma(B)$ and $\kappa(B)$ at low but still finite temperatures
are different, there is a violation of the WF law with observable
oscillations of the Lorenz number \be L(B,T) \equiv
\frac{\kappa(B,T)}{\sigma(B,T) T}. \ee

We begin by presenting in Sec.~\ref{sec:model} the model Lagrangian
describing Dirac quasiparticles in graphene. The general
expressions for electrical and thermal conductivities are written
in terms of the same spectral function $\mathcal{A}(B)$ and the
thermal conductivity is considered also including the thermal
power term. In Sec.~\ref{sec:oscill} we analytically extract
magnetic oscillations contained in the spectral function
$\mathcal{A}(B)$. In the Discussion, Sec.~\ref{sec:disc}, our final
results are summarized and their applicability for the graphite is
considered.

\section{Model and general expressions for electrical
and thermal conductivities}
\label{sec:model}

\subsection{Model Lagrangian}

The Lagrangian density of noninteracting quasiparticles in a
single graphene sheet in the continuum limit reads (see e.g.
\cite{Semenoff:1984:PRL,Gonzales:1993:NP})
\begin{equation}
\label{Lagrangian.Carbon} \mathcal{L}_0 = \sum_{\sigma= \pm 1} v_F
\bar{\Psi}_{\sigma} (t, \mathbf{r}) \left[ \frac{i \gamma^0
(\partial_t - i \mu_\sigma)}{v_F} - i \gamma^1 (\partial_x -ieA_x)
- i \gamma^2 (\partial_y -i eA_y) \right] \Psi_{\sigma}(t,
\mathbf{r}),
\end{equation}
where the four-component Dirac spinor $\Psi_{\sigma} = (\psi_{1
\sigma}, \psi_{2 \sigma})$ is combined from two spinors $\psi_{1
\sigma}$, $\psi_{2 \sigma}$ that describe the Bloch states
residing on the two different sublattices of the biparticle
hexagonal lattice of the graphene sheet, and $\sigma = \pm 1$ is the
spin. In Eq.~(\ref{Lagrangian.Carbon}) $\bar{\Psi}_\sigma =
\Psi_{\sigma}^{\dagger} \gamma^0$ is the Dirac conjugated spinor
and $4\times 4$ $\gamma$ matrices are either $(\sigma_3, i
\sigma_3, - i\sigma_1) \otimes \sigma_3$
\cite{Khveshchenko:2001:PRL} or their unitary-equivalent
representation can be taken from
Ref.~\cite{Semenoff:1984:PRL,Gorbar:2002:PRB}.

Note that there is no principal difference between between
two-band models for electron and holes discussed in
Refs.~\cite{Tokumoto:2004:SSC,Du:2004} and a model with Dirac
fermions, where these electron and holes with identical velocities
are built in the formalism. There are, however, some cases like a
double-resonant Raman scattering in graphite
\cite{Thomsen:2000:PRL}, where the asymmetry between the bonding,
$E = v_F |\mathbf{k}|$ and antibonding $E = - v_F^\prime
|\mathbf{k}|$ bands in graphite is essential, so that multiband
models are more suitable.

Since the terms with $\partial_{x,y}$ in
Eq.~(\ref{Lagrangian.Carbon}) originate from the usual kinetic
term of the tight-binding Hamiltonian, vector potential
$\mathbf{A}$ is inserted in the Lagrangian
(\ref{Lagrangian.Carbon}) using a minimal coupling prescription.
The vector potential for the external magnetic field $\mathbf{B}$
perpendicular to the plane is taken in the symmetric  gauge
\begin{equation}
\label{gauge} \mathbf{A} = \left(-\frac{B}{2}x_2 , \frac{B}{2} x_1
\right).
\end{equation}
Using for the value of the nearest-neighbor hopping matrix element
of graphite $t \sim 2.3 \mbox{eV}$, we obtain that the Fermi
velocity is $v_{F} \approx 7.4 \times 10^5 \mbox{m/s}$, and
accordingly one can estimate from Eq.~(\ref{rel}) that $E_1 \sim
300K \cdot \sqrt{B[\mbox{Tesla}]}$.

Since the Lagrangian (\ref{Lagrangian.Carbon}) originates from
nonrelativistic many-body theory, the interaction of the spin
degree of freedom with magnetic field
\begin{equation}\label{Zeeman}
\mathcal{L}_B =  \mu_B B \sum_{\sigma = \pm} \sigma
\bar{\Psi}_{\sigma} (t, \mathbf{r}) \gamma^0 \Psi_{\sigma}(t,
\mathbf{r})
\end{equation}
has to be explicitly included by considering spin splitting
$\mu_\sigma = \mu - \sigma \mu_B B$ \cite{Shoenberg.book} of the
chemical potential $\mu$, where $\mu_B = e \hbar/(2 m c)$ is the Bohr
magneton. Note that the number of spin components can be regarded
an additional adjustable flavor index of fermions $\sigma =1,
\ldots, N$ and $N=2$ corresponds to the physical case. The
magnitude of the Zeeman term depends on the ratio $\mu_B/k_B
\simeq 0.67 \mbox{K} \cdot \mbox{Tesla}^{-1}$. This term, in fact,
has the same magnitude as the distance between Landau levels in
the nonrelativistic problem. Although we will include this term
for completeness in the analytical expressions, in the numerical
calculations it can be safely neglected because it is much smaller
than estimated above $E_1$.

To make the treatment more general, we also include a mass (gap) term
\begin{equation}\label{gap}
\mathcal{L}_\Delta = \sum_{\sigma= \pm 1} \bar{\Psi}_{\sigma} (t,
\mathbf{r}) \Delta \Psi_{\sigma}(t, \mathbf{r}),
\end{equation}
in the Lagrangian (\ref{Lagrangian.Carbon}). For example, it is
well known that an external magnetic field is a strong catalyst in
generating such a gap for Dirac fermions (the phenomenon of
magnetic catalysis) \cite{Gusynin:1995:PRD}. Usually the opening
of the gap marks an important transition which occurs in the
system. In particular, in the case of pyrolytic graphite a poor
screening of the Coulomb interaction may lead to excitonic
instability, resulting in the opening of the gap in the electronic
spectrum and manifesting itself through the onset of an insulating
charge density wave (see e.g.
\cite{Khveshchenko:2001:PRL,Gorbar:2002:PRB,Kopelevich:2003:PRL}).

\subsection{Electrical conductivity}

The dc conductivity tensor can be found using Kubo formula
\cite{Mahan:book}
\begin{equation}
\label{el-Kubo} \sigma_{ij} = - \lim_{\Omega \to 0}
\frac{\mbox{Im} \Pi^{R}_{ij}(\Omega +i 0)}{\Omega},
\end{equation}
where $\Pi^{R}_{ij}(\Omega)$ is the retarded current-current
correlation function (see e.g.
\cite{Gorbar:2002:PRB,Sharapov:2003:PRB})
\begin{equation}\label{Pi.ij}
\Pi^{R}_{ij}(\Omega + i0) = \frac{e^2 v_F^2}{2} \sum_\sigma
\int_{-\infty}^{\infty} d \omega_1 d \omega_2 \frac{\tanh
[(\omega_2 - \mu_\sigma)/2T] - \tanh
[(\omega_1-\mu_\sigma)/2T]}{\omega_1 - \omega_2 + \Omega + i0}
\int \frac{d^2 k}{(2 \pi)^2} \mbox{tr}[A(\omega_1,\mathbf{k})
\gamma_i A(\omega_2, \mathbf{k}) \gamma_j].
\end{equation}
Here $A(\omega, \mathbf{k})$ is the spectral function associated with
the translationary-invariant part of the Green's function
of Dirac quasiparticles in an external magnetic field
given by Eqs.~(3.3) and (3.6) of
Ref.~\cite{Sharapov:2004:PRB}
(see also Refs.~\cite{Gusynin:1995:PRD,Gorbar:2002:PRB,Sharapov:2003:PRB}).
Then for the diagonal conductivity $\sigma = \sigma_{xx} =
\sigma_{yy}$ we have
\begin{equation}
\lb{sigma}
\sigma = \pi e^2 v_F^2  \sum_\sigma \int  \frac{d^2
k}{(2 \pi)^2} \int \limits_{-\infty}^{\infty} d\omega
[-n_F^{\prime}(\omega - \mu_\sigma)] \mbox{tr} \left[ A(\omega,
{\bf k}) \gamma_1 A(\omega, {\bf k}) \gamma_1 \right] ,
\end{equation}
where $-n_{F}^{\prime}(\omega -\mu) = (1/4T) \cosh^{-2}[(\omega
-\mu)/2T]$ is the derivative of the Fermi distribution. Further
details of calculation of $\mbox{tr}$ and the momentum integral in
Eq.~(\ref{sigma}) were considered in
Refs.~\cite{Ferrer:2003:EPJB,Gorbar:2002:PRB,Sharapov:2003:PRB},
so here we write down a rather simple final expression for the
electrical conductivity in an external magnetic field
\begin{equation}
\lb{sigma-final}
\sigma =  e^2 \sum_\sigma  \int
\limits_{-\infty}^{\infty} \frac{d \omega}{4T\cosh^2 \frac{\omega
- \mu_{\sigma}}{2T}} \mathcal{A}(\omega,B,\Gamma,\Delta),
\end{equation}
where  the function
\begin{equation}
\lb{A.def}
\begin{split}
\mathcal{A}(\omega,B,\Gamma,\Delta)= \frac{1}{\pi^2}
\frac{\Gamma^2}{(e B)^2 + (2 \omega \Gamma)^2}  & \left\{ 2
\omega^2 + \frac{(\omega^2 + \Delta^2 + \Gamma^2)(eB)^2 - 2
\omega^2 (\omega^2 - \Delta^2 + \Gamma^2) eB}
{(\omega^2 - \Delta^2 - \Gamma^2)^2 + 4 \omega^2 \Gamma^2} \right. \\
&\left. - \frac{\omega(\omega^2 - \Delta^2 + \Gamma^2)}{\Gamma}
\mbox{Im} \psi \left( \frac{\Delta^2 + \Gamma^2 - \omega^2 - 2 i
\omega \Gamma} {2eB}\right) \right\}.
\end{split}
\end{equation}
Here in order  to consider the MO for a more realistic case, we
introduced the effect of quasiparticle scattering by making
$\delta$-like quasiparticle peaks associated with the Landau
levels Lorentzians with a constant energy-independent width
$\Gamma$ (see details in Ref.~\cite{Sharapov:2004:PRB}). This
approximation still allows us to derive a rather simple analytical
expression (\ref{A.def}), where $\psi$ is the digamma function, which
eventually results in a {\em Dingle\/} factor in the expression
for the amplitude of MO.

Note that in general one should consider dressed fermion
propagators that include the  self-energy $\Sigma(\omega)$ due to
the scattering from impurities. Up to now the problem of
scattering from impurities in the presence of a magnetic field
does not have yet a satisfactory solution. Therefore, here we have
chosen the case of constant width $\Gamma = \Gamma(\omega = 0) =
-\mbox{Im} \Sigma^R(\omega =0) = 1/(2 \tau)$, $\tau$ being the mean
free time of quasiparticles.

Such a Lorentzian broadening of Landau levels with a constant
$\Gamma$ was found to be a rather good approximation valid in not
very strong magnetic fields \cite{Prange.book,Shoenberg.book}.
Definitely, the treatment of disorder in the presence of the
magnetic field in such a simplified manner should be considered as
only a first step until further progress in this problem is
achieved (in connection with this, see,
Refs.~\cite{Champel:2002:PRB,Grigoriev:2003:PRB}).

\subsection{Thermal conductivity}
\lb{sec:therm}

The longitudinal thermal conductivity can also be calculated using a thermal
Kubo formula \cite{Mahan:book}
\be
\lb{therm-Kubo}
\frac{\kappa(B,T)}{T} = - \frac{1}{T^2} \lim_{\Omega \to 0}
\frac{\mbox{Im} \Pi_{EE}^R (\Omega +i0)}{\Omega} - \frac{1}{T^2 \sigma}
\lim_{\Omega \to 0} \frac{[\mbox{Im} \Pi_{EC}^{R}(\Omega +i0)]^2}{\Omega^2}.
\ee
Here $\Pi_{EE}^{R}(\Omega)$ is the retarded longitudinal
energy current-current correlation function and $\Pi_{EC}^R(\Omega)$
is the retarded longitudinal correlation function of energy current
with electrical current and the expression for
the energy current operator is given
in Refs.~\cite{Ferrer:2003:EPJB,Sharapov:2003:PRB}.

We note in passing that the second term of Eq.~(\ref{therm-Kubo})
is related to the thermal power \be \lb{thermal-power} S = -
\frac{1}{T} \lim_{\Omega \to 0} \frac{\mbox{Im}
\Pi_{EC}^{R}(\Omega +i0)}{\mbox{Im} \Pi^R (\Omega + i0)}, \ee
where $\Pi^R (\Omega + i0) = \Pi^R_{xx} (\Omega + i0) = \Pi^R_{yy}
(\Omega + i0)$. The presence of the thermal power term in
Eq.~(\ref{therm-Kubo}) ensures that the energy current is
evaluated under the condition of vanishing electrical current
\cite{Mahan:book}.

Similarly to the above-derived  electrical conductivity we finally
arrive at
\begin{equation}
\lb{therm-cond-final}
\begin{split}
\frac{\kappa (B,T)}{T} =  &  \sum_\sigma \int
\limits_{-\infty}^{\infty} d \omega \left(
\frac{\omega-\mu_\sigma}{T} \right)^2 \frac{1}{4T\cosh^2
\frac{\omega - \mu_\sigma}{2T}}
\mathcal{A}(\omega,B,\Gamma,\Delta)  \\
& -  \frac{e^2 }{\sigma (B,T)} \left[ \sum_\sigma \int
\limits_{-\infty}^{\infty} d \omega  \frac{\omega-\mu_\sigma}{T}
\frac{1}{4T\cosh^2 \frac{\omega - \mu_\sigma}{2T}}
\mathcal{A}(\omega,B,\Gamma,\Delta) \right]^2  ,
\end{split}
\end{equation}
where $\mathcal{A}$ is the same function (\ref{A.def}) as for the
electrical conductivity. This function contains all information
about the field dependence of the transport properties of the
systems with a linear dispersion law, including the MO. While the
representation (\ref{A.def}) can already be used for numerical
calculations, for analytical work it is useful to extract
explicitly the MO that are contained in the digamma function
$\psi$ when the real part of its argument becomes negative.

\section{Analytical consideration of oscillations}
\lb{sec:oscill}

\subsection{Extracting oscillations from $\mathcal{A}$ using
$\psi$-function properties}

 The oscillations of
$\mathcal{A}$ in $1/B$ can be extracted using the relationship for
$\psi$ function
\begin{equation}\label{psi-osc}
\psi(-z)=\psi(z)+\frac{1}{z} + \pi \cot \pi z,
\end{equation}
which [see also Eq.~(4.18) of Ref.~\cite{Sharapov:2004:PRB}]
results in the expression 
\be
\begin{split}
&\psi\left(\frac{\Delta^2-(\epsilon+i\Gamma)^2}{2eB}\right)=
\psi\left(\frac{\Delta^2-(\epsilon+i\Gamma)^2}{2eB}\right)
\left[\theta(\epsilon^2-\Delta^2-\Gamma^2)+\theta(\Delta^2+\Gamma^2-
\epsilon^2)\right]\\
&={\rm
Re}\psi\left(\frac{|\epsilon^2-\Delta^2-\Gamma^2|-2i\epsilon\Gamma}
{2eB}\right)-i\,{\rm sgn}(\epsilon^2-\Delta^2-\Gamma^2){\rm
Im}\psi\left(
\frac{|\epsilon^2-\Delta^2-\Gamma^2|-2i\epsilon\Gamma}{2eB}\right)\\
&+\theta(\epsilon^2-\Delta^2-\Gamma^2)\left[\frac{2eB}{\epsilon^2
-\Delta^2-\Gamma^2+2i\epsilon\Gamma}+\pi\cot\pi\frac{\epsilon^2
-\Delta^2-\Gamma^2+2i\epsilon\Gamma}{2eB}\right].
\end{split}
\ee 
Taking the imaginary part of the last equation, we obtain 
\be
\label{Im.psi}
\begin{split}
& \mbox{Im}
\psi\left(\frac{\Delta^2-(\omega+i\Gamma)^2}{2eB}\right)=  -
\mbox{sgn} (\omega^2 - \Delta^2 - \Gamma^2)
\mbox{Im}\psi\left(\frac{|\omega^2 - \Delta^2 - \Gamma^2|- 2 i\omega \Gamma}{2eB}\right) \\
& - \theta (\omega^2 - \Delta^2 - \Gamma^2) \left[\frac{4 eB
\omega \Gamma}{(\omega^2 - \Delta^2 - \Gamma^2)^2 + 4 \omega^2
\Gamma^2} + \pi \frac{\sinh (2 \pi \omega \Gamma/eB)}{\cosh (2 \pi
\omega \Gamma/eB) - \cos [\pi(\omega^2 - \Delta^2 -
\Gamma^2)/eB]}\right].
\end{split}
\ee Then substituting Eq.~(\ref{Im.psi}) in Eq.~(\ref{A.def}) we
obtain
\begin{equation}
\label{A.osc}
\begin{split}
\mathcal{A}& (\omega,B,\Gamma,\Delta) =  \frac{1}{\pi^2}
\frac{\Gamma^2}{(e B)^2 + (2 \omega \Gamma)^2}   \left\{ 2
\omega^2 + \frac{(eB)^2/2} {(\omega+\Delta)^2 + \Gamma^2} +
\frac{(eB)^2/2} {(\omega-\Delta)^2 + \Gamma^2} - \frac{2 \omega^2
(\omega^2 - \Delta^2 + \Gamma^2) eB} {(\omega^2 - \Delta^2 -
\Gamma^2)^2 + 4 \omega^2 \Gamma^2}
\right. \\
& + \frac{\omega(\omega^2 - \Delta^2 + \Gamma^2)}{\Gamma} \left[
\mbox{sgn} (\omega^2 - \Delta^2 - \Gamma^2)
\mbox{Im}\psi\left(\frac{|\omega^2 - \Delta^2 - \Gamma^2|- 2
i\omega \Gamma}{2eB}\right) \right.\\
& \left. \left. + \theta (\omega^2 - \Delta^2 - \Gamma^2)
\left(\frac{4 eB \omega \Gamma}{(\omega^2 - \Delta^2 - \Gamma^2)^2
+ 4 \omega^2 \Gamma^2} + \pi \frac{\sinh (2 \pi \omega
\Gamma/eB)}{\cosh (2 \pi \omega \Gamma/eB) - \cos [\pi(\omega^2 -
\Delta^2 - \Gamma^2)/eB]}\right) \right] \right\},
\end{split}
\end{equation}
where the oscillations are contained in the last term of
Eq.~(\ref{A.osc}). Note that the real part of the argument of
$\psi$ function in (\ref{A.osc}) is already positive and the signs
before $\psi$ in (\ref{A.def}) and (\ref{A.osc}) are different.

For $\omega ^2 > \Delta^2 + \Gamma^2$ using the relationship
\begin{eqnarray}
{\rm Re}\frac{e^{-(a-ib)}}{1-e^{-(a-ib)}}=\frac{\cos
b-e^{-a}}{2(\cosh a-\cos b)},
\end{eqnarray}
one can expand
\begin{equation}\label{series}
\frac{\sinh (2 \pi |\omega| \Gamma/eB)}{\cosh (2 \pi \omega
\Gamma/eB) - \cos [\pi(\omega^2 - \Delta^2 - \Gamma^2)/eB]} = 1 +
2 \sum_{k=1}^{\infty} \cos [\pi k(\omega^2 - \Delta^2 -
\Gamma^2)/eB] \exp (- 2 \pi k |\omega| \Gamma/eB)
\end{equation}
and finally arrive at the expression for oscillatory part of
$\mathcal{A}$
\begin{equation}\label{A.osc.final}
\begin{split}
\mathcal{A}_{\mathrm{osc}}(\omega,B,\Gamma,\Delta) = & \frac{2}{
\pi}\frac{\omega \Gamma (\omega^2 - \Delta^2 + \Gamma^2) \theta (\omega^2 - \Delta^2 - \Gamma^2)}
{(eB)^2 + (2 \omega \Gamma)^2} \\
& \times \sum_{k=1}^{\infty} \cos \frac{\pi k (\omega^2 - \Delta^2
- \Gamma^2)}{eB} \exp \left(-\frac{2 \pi k |\omega| \Gamma}{eB}
\right) .
\end{split}
\end{equation}

\subsubsection{Low-field non oscillatory limit}

Eq.~(\ref{A.osc}) can be simplified in the low field limit:
\begin{equation}\label{A.Bto0.osc}
\begin{split}
\mathcal{A}(\omega,B,\Gamma,\Delta) & =\frac{1}{2 \pi^2} \left[ 1
- \frac{\omega^2 - \Delta^2 + \Gamma^2}{2 |\omega| \Gamma}
\mbox{sign} (\omega^2 -\Delta^2 - \Gamma^2)
\arctan \frac{2 |\omega| \Gamma}{|\omega^2 - \Delta^2 - \Gamma^2|} \right.\\
& \left. + \theta(\omega^2 - \Delta^2 - \Gamma^2)
\frac{\pi(\omega^2 - \Delta^2 + \Gamma^2)}{2 \omega \Gamma}
\frac{\sinh (2 \pi \omega \Gamma/eB)}{\cosh (2 \pi \omega
\Gamma/eB) - \cos [\pi(\omega^2 - \Delta^2 - \Gamma^2)/eB]},
\right],
\end{split}
\end{equation}
where we kept $B$ only in the oscillatory part of $\mathcal{A}$.
For $\omega ^2 < \Delta^2 + \Gamma^2$ after using the relationship
\begin{equation}\label{arctan}
\arctan x = \frac{\pi}{2} - \arctan \frac{1}{x}, \qquad x > 0,
\end{equation}
the last equation reduces to [see also Eq.~(4.16) of
Ref.~\cite{Sharapov:2003:PRB}]
\begin{equation}
\label{A.B=0} \mathcal{A}(\omega,B=0,\Gamma,\Delta)= \frac{1}{2
\pi^2} \left[ 1 + \frac{\omega^2 - \Delta^2 + \Gamma^2}{2 |\omega|
\Gamma} \left(\frac{\pi}{2} - \arctan \frac{\Delta^2 + \Gamma^2 -
\omega^2} {2|\omega| \Gamma} \right) \right].
\end{equation}

\subsection{Oscillating parts of electrical and thermal conductivities}

Substituting Eq.~(\ref{A.osc.final}) in (\ref{sigma-final}) one
can obtain the expression for oscillating part of conductivity
\begin{equation}
\lb{sigma.osc.final}
\begin{split}
\sigma_{\mathrm{osc}} = & \frac{4 e^2 |\mu |\Gamma}{\pi} \frac{
(\mu^2 - \Delta^2 + \Gamma^2) \theta(\mu^2 - \Delta^2 - \Gamma^2)}{(eB)^2 +(2 \mu \Gamma)^2}\\
&\times \sum_{k=1}^{\infty} \cos\left[\frac{\pi k(\mu^2-\Delta^2 -
\Gamma^2)}{eB} \right] R_{T}(k,\mu) R_{D}(k,\mu),
\end{split}
\end{equation}
where we introduced the {\em temperature amplitude\/} factor
\begin{equation}
\lb{temperature-factor} R_{T}(k,\mu) \equiv R_{T} (t_k)=
\frac{t_k}{\sinh t_k}, \qquad t_k = \frac{2\pi^2 k T \mu}{eB}
\qquad (R_{T}(0)=1),
\end{equation}
and the {\em Dingle factor\/}
\begin{equation}
\lb{Dingle-factor} R_{D}(k,\mu) = \exp \left[-\frac{2 \pi k |\mu|
\Gamma}{eB} \right].
\end{equation}
Deriving Eq.~(\ref{sigma.osc.final}) we made the following
simplifying assumptions:  (i) Spin splitting is not included, and (ii)
the low temperature $T\to 0$ limit is considered. Thus after
making a shift $\omega \to \omega + \mu$ and changing the variable
$\omega \to 2 T \omega$, and keeping only the linear in $T$ terms
in the oscillating part of the integrand, we used the integral
\begin{equation}
\lb{integral} \int_{0}^{\infty} d x \frac{\cos b x}{\cosh^2 x} =
\frac{\pi b/2}{\sinh \pi b/2}
\end{equation}
to obtain the temperature amplitude factor
(\ref{temperature-factor}). It is essential that in contrast to
the the Dingle and temperature factors for nonrelativisitic
spectrum, both (\ref{temperature-factor}) and
(\ref{Dingle-factor}) factors for the relativistic spectrum
contain chemical potential $\mu$ (see also
Ref.~\cite{Sharapov:2004:PRB}). The distinctive concentration
dependence of $R_T$ should be observed experimentally.

\begin{figure}[h]
\centering{
\includegraphics[width=8cm]{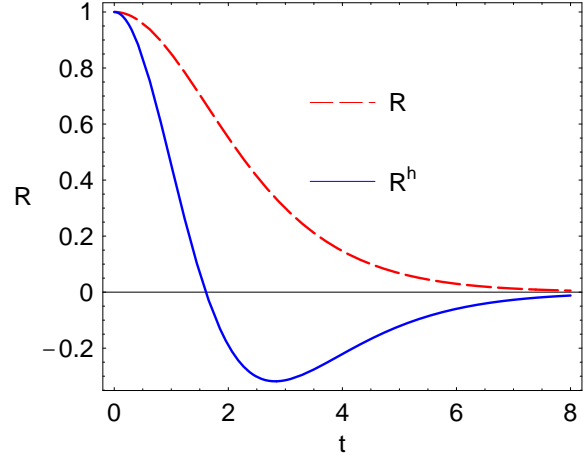}}
\caption{(Color online) The temperature amplitude factors $R_T(t)$
and $R_{T}^h(t)$  given by Eqs.~(\ref{temperature-factor}) and
(\ref{temperature-factor.h}).} \label{fig:1}
\end{figure}

Similarly to Eq.~(\ref{sigma.osc.final}) one arrives at the
expression for the oscillating part of thermal conductivity
(\ref{therm-cond-final}):
\begin{equation}
\lb{kappa.osc.final}
\begin{split}
\frac{\kappa_{\mathrm{osc}}}{T} = & \frac{4 \pi |\mu| \Gamma}{3}
\frac{
(\mu^2 - \Delta^2 + \Gamma^2) \theta(\mu^2 - \Delta^2 - \Gamma^2)}{(eB)^2 +(2 \mu \Gamma)^2}\\
&\times \sum_{k=1}^{\infty} \cos\left[\frac{\pi k(\mu^2-\Delta^2 -
\Gamma^2)}{eB} \right] R_{T}^{h}(k,\mu) R_{D}(k,\mu),
\end{split}
\end{equation}
where
\begin{equation}
\lb{temperature-factor.h} R_{T}^{h}(t_k) =
\frac{12}{\pi^2}\frac{d^2 R_T(t_k)}{d t_k^2}= \frac{6}{\sinh
t_k}\left[ \coth t_k - \frac{t_k}{2} - \frac{t_k}{\sinh^2
t_k}\right], \qquad [R_{T}^h(0)=1]
\end{equation}
is the {\it temperature amplitude\/} factor for thermal
conductivity obtained using the second derivative of the integral
(\ref{integral}). The temperature amplitude factors $R_T(t)$ and
$R_T^h(t)$ are shown in Fig.~\ref{fig:1}. Interestingly the
dependence $R_T^h(t)$ is non-monotonic and $R_T^h(t)$ even changes
its sign.

\section{Discussion}
\lb{sec:disc}

\begin{figure}[h]
\centering{
\includegraphics[width=8cm]{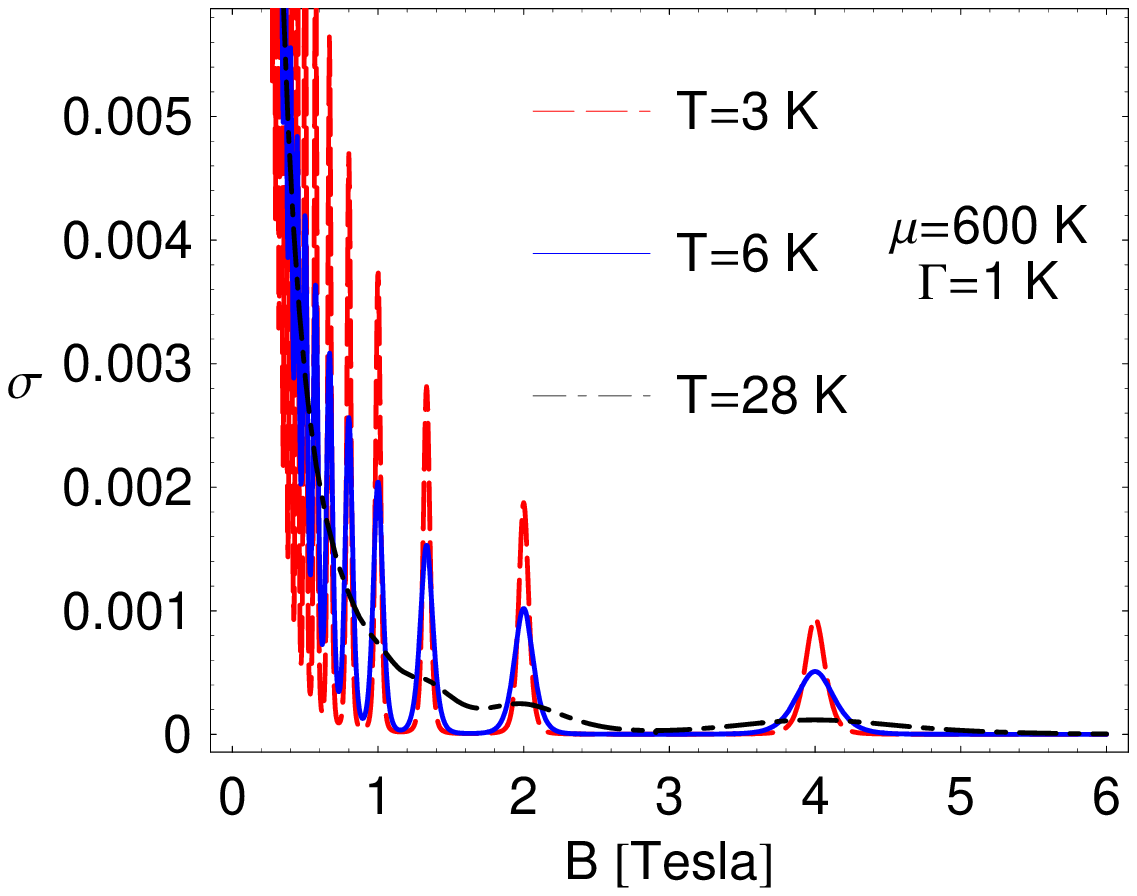}}
\caption{(Color online) The normalized conductivity
$\sigma(B,T)/\sigma(B =0,T)$ as a function of field $B$ for three
different values of temperature $T$ for $\mu= 600 \mbox{K}$ and
$\Gamma= 1 \mbox{K}$. We use $e B \to (4.5 \times 10^4 \mbox{K}^2)
B(\mbox{Tesla})$.} \label{fig:2}
\end{figure}

\begin{figure}[h]
\centering{
\includegraphics[width=8cm]{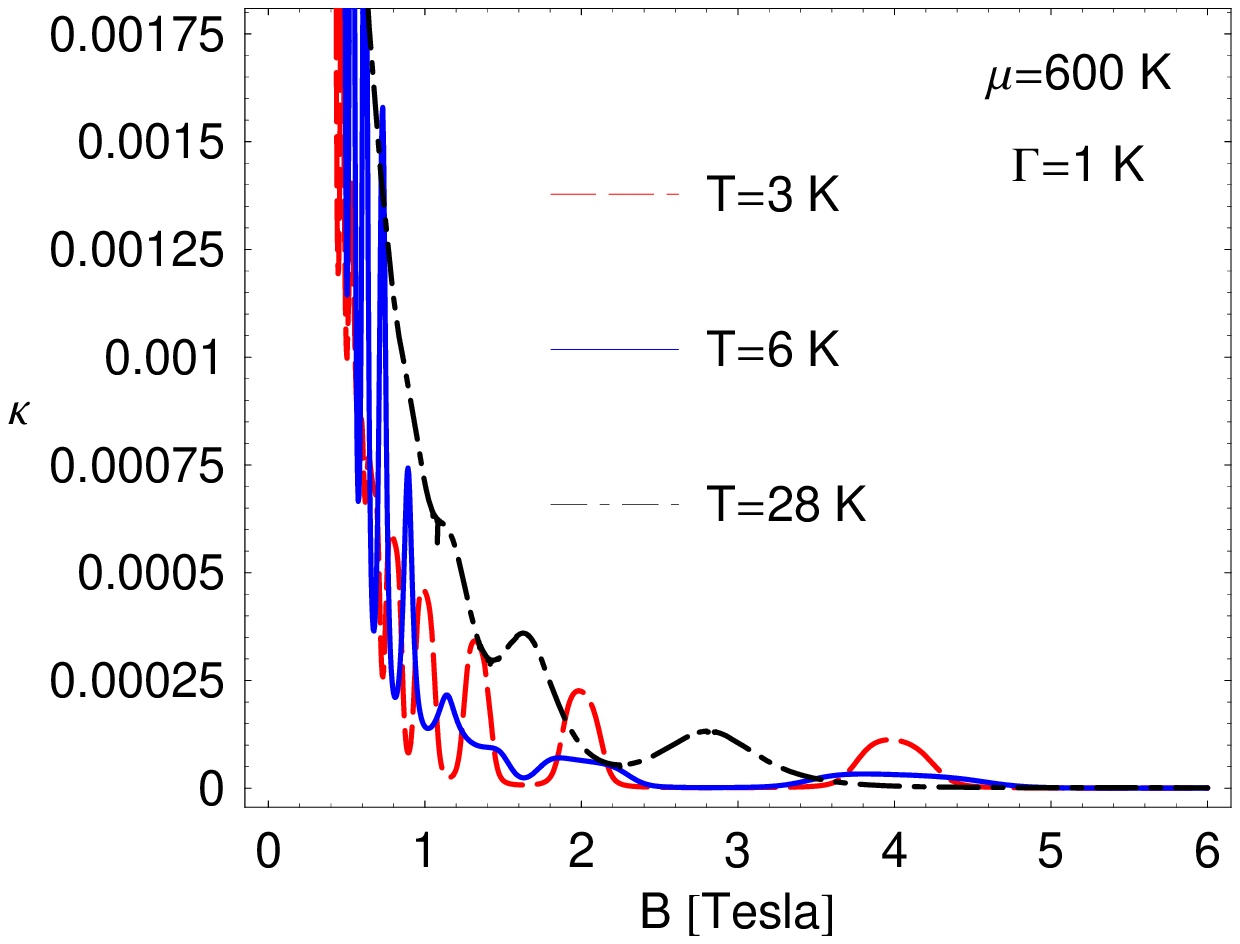}}
\caption{(Color online) The normalized thermal conductivity
$\kappa(B,T)/\kappa(B=0, T)$ as a function of field $B$ for three
different values of temperature $T$ for $\mu= 600 \mbox{K}$ and
$\Gamma= 1 \mbox{K}$. We use $e B \to (4.5 \times 10^4 \mbox{K}^2)
B(\mbox{Tesla})$.} \label{fig:3}
\end{figure}

\begin{figure}[h]
\centering{
\includegraphics[width=8cm]{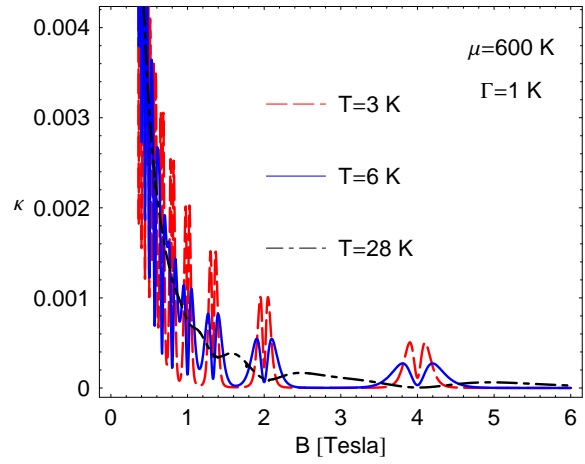}}
\caption{(Color online) The normalized thermal conductivity
$\kappa_0(B,T)/\kappa(B=0, T)$ [calculated without the second term
of Eq.~(\ref{therm-Kubo}] that originates from the condition of
absence electrical current in the system) as a function of field,
$B$, for three different values of temperature $T$ for $\mu= 600
\mbox{K}$ and $\Gamma= 1 \mbox{K}$. We use $e B \to (4.5 \times
10^4 \mbox{K}^2) B(\mbox{Tesla})$.} \label{fig:4}
\end{figure}

\begin{figure}[h]
\centering{
\includegraphics[width=8cm]{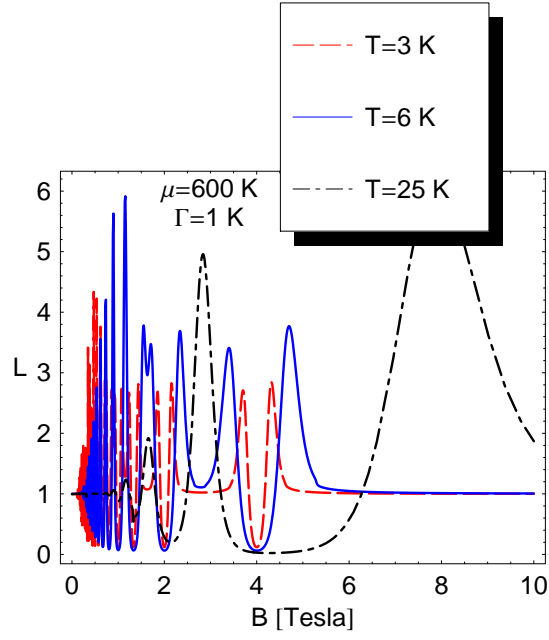}}
\caption{(Color online) The normalized Lorenz number $L(B,T)/L_0$
as a function of field $B$ for three different values of
temperature $T$ for $\mu= 600 \mbox{K}$ and $\Gamma= 1 \mbox{K}$.
We use $e B \to (4.5 \times 10^4 \mbox{K}^2) B(\mbox{Tesla})$.}
\label{fig:5}
\end{figure}

Based on Eqs.~(\ref{sigma-final}) and (\ref{therm-cond-final}),
in Figs.~\ref{fig:2}--\ref{fig:5} we compute the field dependences
of $\sigma(B,T)/\sigma(B=0,T)$, and $\kappa(B,T)/\kappa(B=0,T)$ and
the normalized Lorenz number $L(B,T)/L_0 $, where $L_0 = \pi^2
k_B^2/(3 e^2)$ is the Sommerfeld's value for the Lorenz ratio. To
simplify our consideration we set $\Delta =0$, but it may  be
necessary and quite interesting to consider the influence of
$\Delta$ on MO; see Ref.~\cite{Sharapov:2004:PRB}. As mentioned
above Eq.~(\ref{gap}), we also do not include spin splitting.
Analyzing these figures one can conclude the following.

\begin{enumerate}

\item As usually, the conditions favorable for the magnetic
oscillations are $\Gamma, T \ll \omega_L$, where $\omega_L \sim E_1$
is the distance between Landau levels. This regime is different
from the regime $\Gamma \ll \omega_L \ll T$ of an unconventional
magnetotransport \cite{Du:2004} when MO's are still not resolved
due to the thermal smearing of Landau levels.

\item As the field increases from $B=0$ both electrical and thermal
conductivities decrease rapidly and start to oscillate when Landau
levels cross the Fermi surface. The specific of graphene is that
for the realistic values of the parameters there are only a few
Landau levels below the Fermi surface. As the field $B$ increases,
these levels quickly cross the Fermi surface, and as one can see
from Fig.~\ref{fig:2} for $B > 4 \mbox{Tesla}$ the lowest Landau
level $E_1$ is already above the Fermi surface so that the MO's
disappear and the system can, in principle, enter in the quantum
Hall effect regime. Thus MO's in graphene with $e B \lesssim \mu^2$
are quite different from the conventional MO's when there are so
many Landau levels below the Fermi surface that $\omega_L \ll \mu$
and when a new level crosses the Fermi surface the system returns
practically in the same state as before.

For Dirac fermions in order to have at least one oscillation the
inequality [see Eq.~(8.20) of \cite{Sharapov:2004:PRB}] $\mu^2 -
\Delta^2 \geq 2 e B$ have to be satisfied, and analyzing the
experimental data of Refs.~\cite{Luk'yanchuk:2004,Ocana:2003:PRB}
there is a temptation to distinguish there the regimes of
conventional MO's and quantum Hall effect. However, the real
situation is more involved due to the difference between the model
for graphene with linearized spectrum and measured properties of
bulk graphite (see also Ref.~\cite{Matsui:2004}). Nevertheless, the
observation of plateaulike  features in the Hall
resistivity for $B \gtrsim 2 \mbox{Tesla}$ \cite{Ocana:2003:PRB}
suggests that for these fields the system is already in the
quantum Hall effect regime.

\item Comparing Figs.~\ref{fig:2} and \ref{fig:3}, one can see
that $\sigma(B)$ and $\kappa(B)$ do not oscillate in phase. In
particular, we observe that at $T = 28 \mbox{K}$ the oscillations
of $\sigma(B)$ are practically invisible, while one can still
notice some oscillations of $\kappa(B)$.

To look closer at the difference between $\sigma$ and $\kappa$ in
Fig.~\ref{fig:4} we also plotted the thermal conductivity
$\kappa_0(B,T)$ calculated without the second term of
Eq.~(\ref{therm-Kubo}). In normal metals this second term is
considered to be unimportant because usually it is $\sim
T^2/\mu^2$ times less than the first term of
Eq.~(\ref{therm-Kubo}) and because the WF law is always
considered in the limit $T \ll \mu$  this term is usually
neglected.

\item
By comparing Fig.~\ref{fig:2} and Fig.~\ref{fig:4},  it is easy to
see that each peak of the electrical conductivity is accompanied
by two satellite peaks of the thermal conductivity. The dip
between these two peaks in $\kappa(B)$ coincides with the peak of
$\sigma(B)$. The origin of these satellite peaks is related to the
fact that the expression for the thermal conductivity
(\ref{therm-Kubo}) contains the factor $g(\omega) = -
n_F^{\prime}(\omega - \mu) (\omega- \mu)^2/T^2$ and thus measures
$\mathcal{A}(\omega, B, \Gamma)$ below and above the Fermi energy,
while the electrical conductivity probes $\mathcal{A}(\omega, B,
\Gamma)$ {\em at} the Fermi energy, because it contains just the
factor $n_F^\prime(\omega)$ (see \cite{Taylor.book} and Figs.~12
and 13 of Ref.~\cite{Sharapov:2003:PRB}).

\item Let us now  compare Figs.~\ref{fig:3}
and \ref{fig:4}. At $T = 3  \mbox{K}$ each double-peak structure
observed in Fig.~\ref{fig:4} is replaced by a single broader peak
in Fig.~\ref{fig:3}. This reflects the fact that the full
expression (\ref{therm-Kubo}) for $\kappa(B) \sim \kappa_0(B) -
a(B)/\sigma(B)$, so that if the coefficient $a(B)$ is  large
enough, the peaks seen in $\sigma(B)$ also produce an increase of
$\kappa(B)$, so the dip between peaks in $\kappa(B)$ is filled in
and we observe a single broader peak.

For higher temperatures the role played by second term of
Eq.~(\ref{therm-Kubo}) further increases and $\kappa(B)$ and
$\kappa_0(B)$ behave quite differently. Finally we observe the
above-mentioned  picture when at $T = 28 \mbox{K}$ the oscillations
of $\sigma(B)$ are damped, but one can still see some oscillations
of $\kappa(B)$, but the positions of the peaks do not coincide
with the positions of the peaks in $\sigma(B)$ and $\kappa(B)$
observed at lower temperatures.

\item The behavior of Lorenz
number is shown in Fig.~\ref{fig:5}. Since in the chosen
temperature interval the inequality $T \ll \mu$ is well justified,
we observe that the WF law is maintained in zero field (see
Ref.~\cite{Sharapov:2003:PRB}). In nonzero field we observe
violations of the WF law that become more strong as the
temperature increases. At $T = 3 \mbox{K}$ the behavior of $L(B)$
is similar to the behavior of $\kappa_0(B)$ and two satellite
peaks in $L(B)$ are related to the broad peak in $\kappa(B)$,
while the dip between these peaks is caused by the peak of
$\sigma(B)$. As the temperature increases a more complicated
behavior of $\kappa(B)$ results in the large-amplitude
oscillations of $L(B)$. The positions of the peaks are not related
to the positions of low-$T$ SdH oscillations. In the high-field
regime $B > 10 \mbox{Tesla}$ there is a tendency to the
restoration of the WF law.

\end{enumerate}

The above-presented  picture is already quite complicated due to
the interplay between the first and second terms of
Eq.~(\ref{therm-Kubo}), so that in general there are no
correlations between the low temperature SdH oscillations and
oscillations in $\kappa(B)$ and $L(B)$ seen at higher
temperatures. Further complications can be caused by the fact that
the impurity scattering rate $\Gamma$ which we assumed to be field
and temperature independent may, in fact, depend on both $B$ and $T$
\cite{Champel:2002:PRB,Grigoriev:2003:PRB}.

Let us now discuss the relation of the obtained theoretical
results to the experiments
\cite{Luk'yanchuk:2004,Ocana:2003:PRB,Ulrich:2004:JLTP}. To
compare the results for electrical conductivity shown in
Fig.~\ref{fig:2}, one should bear in mind the difference between
graphene and graphite mentioned in item 2. In
Ref.~\cite{Luk'yanchuk:2004} the measurements were done at $T = 2
\mbox{K}$ and the oscillations are clearly seen and some of them,
as stated in Ref.~\cite{Luk'yanchuk:2004}, are related to the
quasiparticles with a linear dispersion. It is likely that for
higher temperatures the oscillations of $\sigma(B)$ are suppressed
\cite{Ocana:2003:PRB} and we also see that at $T = 28 \mbox{K}$
the oscillations of $\sigma(B)$ are practically smeared out.

A comparison with experimental results for thermal conductivity is
more complicated because the measured thermal conductivity
$\kappa_{\mathrm{exp}}(B,T)=
\kappa(B,T)+\kappa_{\mathrm{ph}}(B,T)$ besides the electron
contribution, $\kappa(B,T)$ contains also the contribution from
phonons, $\kappa_{\mathrm{ph}}(T)$, which is assumed to be field
independent. Accordingly, we can relate theoretically calculated
$\kappa(B,T)/\kappa(B=0,T)$ shown in Fig.~\ref{fig:3} with the
experimentally accessible ratio
$\kappa_{\mathrm{exp}}(B,T)/\kappa_{\mathrm{exp}}(B=0,T)$ via
\begin{equation}
\lb{kappa.exp}
\frac{\kappa_{\mathrm{exp}}(B,T)}{\kappa_{\mathrm{exp}}(B=0,T)} =
1+
\frac{\kappa(B,T)/\kappa(B=0,T)-1}{\kappa_{\mathrm{ph}}(T)/\kappa(B=0,T)+1}
\approx 1+ \frac{\kappa(B,T)/\kappa(B=0,T)-1}{(T/T_0)^2+1}.
\end{equation}
The last equality is written using that the ratio
$\kappa_{\mathrm{ph}}(T)/\kappa(B=0,T)$ can be determined from the
fact that at $T = T_0 = 1.5 \mbox{K}$ the values
$\kappa_{\mathrm{ph}}(T_0) \sim \kappa(B=0,T_0)$
\cite{Ulrich:2004:JLTP}, so that assuming $\kappa_{\mathrm{ph}}(T)
\sim T^3$ and $\kappa(B=0,T) \sim T$ we estimate
$\kappa_{\mathrm{ph}}(T)/\kappa(B=0,T) \sim (T/T_0)^2$.

\begin{figure}[h] \centering{
\includegraphics[width=8cm]{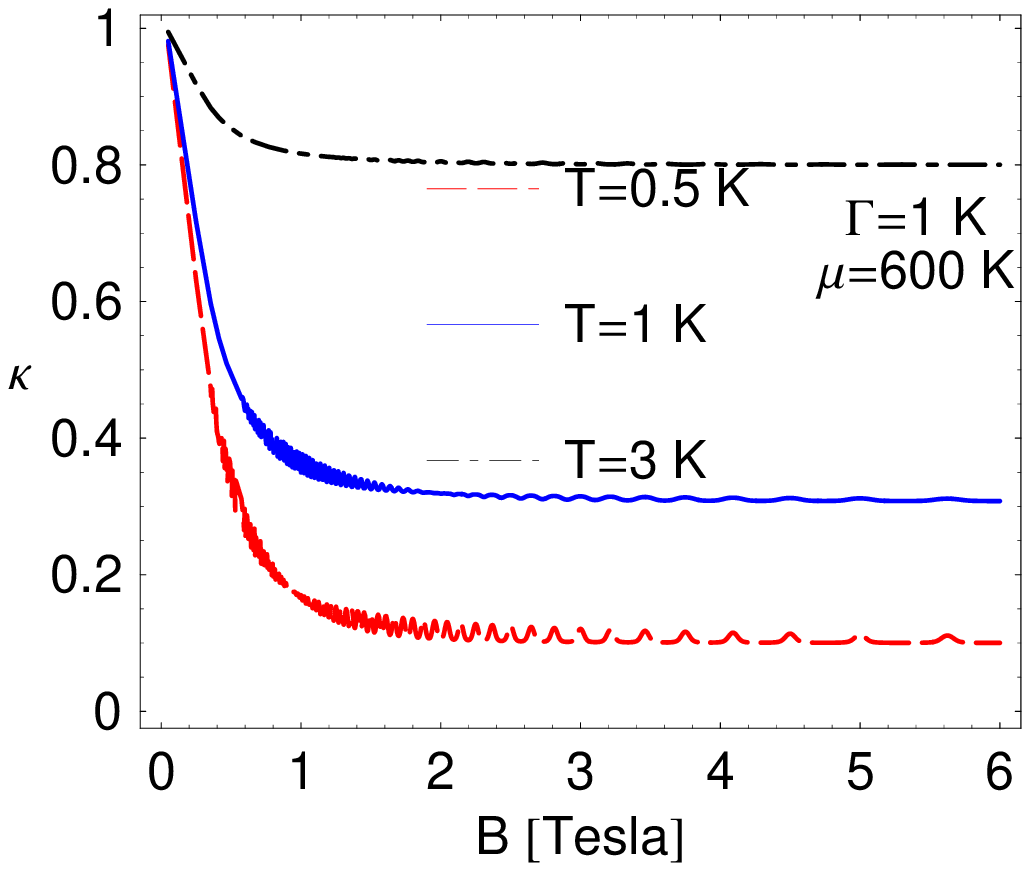}}
\caption{(Color online) The normalized thermal conductivity
$\kappa_{\mathrm{exp}}(B,T)/\kappa_{\mathrm{exp}}(B=0, T)$ [see
Eq.~(\ref{kappa.exp})] as a function of field $B$ for three
different values of temperature $T$ for $\mu= 600 \mbox{K}$ and
$\Gamma= 1 \mbox{K}$. We use $e B \to (4.5 \times 10^4 \mbox{K}^2)
B(\mbox{Tesla})$ and $T_0=1.5 \mbox{K}$.} \label{fig:6}
\end{figure}

The ratio
$\kappa_{\mathrm{exp}}(B,T)/\kappa_{\mathrm{exp}}(B=0,T)$ is
plotted in Fig.~\ref{fig:6}. As one can see the MO's  of
$\kappa(B,T)$ are masked by the phonon contribution in
$\kappa_{\mathrm{exp}}(B,T)$ and only at rather low temperatures
is there  a possibility to observe them. Moreover, since the
amplitude of MO is higher for lower fields (see also
Figs.~\ref{fig:2}-\ref{fig:4}), the MO's are more
easily observable in the low-field regime.

Taking these facts into account let us discuss the oscillations of
thermal conductivity that were experimentally observed in
Refs.~\cite{Ocana:2003:PRB,Ulrich:2004:JLTP} for $B \gtrsim 2
\mbox{Tesla}$. Since these oscillations are clearly seen only in
the high field regime and their amplitude increases as the field
grows, their origin is unlikely related to the conventional
SdH-like MO's and seems to be more associated to the quantum Hall
effect. It would be interesting to check whether the conventional
MO's of the electronic thermal conductivity $\kappa(B,T)$ predicted
in this paper can be also observed in graphite when the
measurements will be done at sufficiently low temperatures and/or
the phonon contribution is subtracted.

\section{Acknowledgments}
We gratefully acknowledge P.~Esquinazi and  A.~Geim for a helpful
discussions.

\end{document}